\documentclass[sigconf,10pt]{acmart}

\usepackage{booktabs} 
\usepackage{todonotes}
\usepackage{xcolor,listings}
\usepackage{lipsum}
\usepackage{amssymb}
\usepackage{tikz}
\usetikzlibrary{calc} 
\usetikzlibrary{shapes,snakes} 
\usetikzlibrary{decorations.pathreplacing}
\usetikzlibrary{patterns}

\usepackage{enumitem} 

\usepackage{comment}

\usepackage{multirow}

\usepackage[flushleft]{threeparttable}

\copyrightyear{2019} 
\acmYear{2019} 
\setcopyright{licensedusgovmixed}
\acmConference[SIGMOD '19]{2019 International Conference on Management of Data}{June 30-July 5, 2019}{Amsterdam, Netherlands}
\acmBooktitle{2019 International Conference on Management of Data (SIGMOD '19), June 30-July 5, 2019, Amsterdam, Netherlands}
\acmPrice{15.00}
\acmDOI{10.1145/3299869.3314040}
\acmISBN{978-1-4503-5643-5/19/06}

\newtheorem{extension}{Extension}

\begin{document}
\title[One SQL to Rule Them All]{One SQL to Rule Them All:\\ An Efficient and Syntactically Idiomatic Approach to Management of Streams and Tables}

\subtitle{An Industrial Paper}

\author{Edmon Begoli}
\orcid{0000-0002-2173-3663}
\affiliation{%
  \institution{Oak Ridge National Laboratory / Apache Calcite}
  \streetaddress{1 Bethel Valley Rd.}
  \city{Oak Ridge}
  \state{Tennessee}
  \country{USA}
  \postcode{37831}
}
\email{begoli@apache.org}

\author{Tyler Akidau}
\affiliation{%
  \institution{Google Inc. / Apache Beam}
  \streetaddress{601 N 34th St}
  \city{Seattle}
  \state{WA}
  \country{USA}
  \postcode{98103}
}
\email{takidau@apache.org}

\author{Fabian Hueske}
\affiliation{%
  \institution{Ververica / Apache Flink}
  \streetaddress{Stresemannstraße 121A}
  \city{Berlin}
  \country{Germany}
  \postcode{10963}
}
\email{fhueske@apache.org}

\author{Julian Hyde}
\affiliation{%
  \institution{Looker Inc. / Apache Calcite}
  \streetaddress{2300 Harrison Street}
  \city{San Francisco}
  \state{California}
  \country{USA}
  \postcode{94110}
}
\email{jhyde@apache.org}

\author{Kathryn Knight}

\affiliation{%
  \institution{Oak Ridge National Laboratory}
  \streetaddress{1 Bethel Valley Rd.}
  \city{Oak Ridge}
  \state{Tennessee}
  \country{USA}
  \postcode{37831}
}
\email{knightke@ornl.gov}

\author{Kenneth Knowles}
\affiliation{%
  \institution{Google Inc. / Apache Beam}
  \streetaddress{601 N 34th St}
  \city{Seattle}
  \state{WA}
  \country{USA}
  \postcode{98103}
}
\email{kenn@apache.org}

\renewcommand{\shortauthors}{Begoli et al.}


\begin{abstract}

Real-time data analysis and management are increasingly critical for today's businesses. SQL is the de facto \textit{lingua franca} for these endeavors, yet support for robust streaming analysis and management with SQL remains limited. Many approaches restrict semantics to a reduced subset of features and/or require a suite of non-standard constructs. Additionally, use of event timestamps to provide native support for analyzing events according to when they actually occurred is not pervasive, and often comes with important limitations.


We present a three-part proposal for integrating robust streaming into the SQL standard, namely: (1) time-varying relations as a foundation for classical tables as well as streaming data, (2) event time semantics, (3) a limited set of optional keyword extensions to control the materialization of time-varying query results. Motivated and illustrated using examples and lessons learned from implementations in Apache Calcite, Apache Flink, and Apache Beam, we show how with these minimal additions it is possible to utilize the complete suite of standard SQL semantics to perform robust stream processing.


\end{abstract}

%
%



\maketitle

\newcommand{\istream}{\ensuremath{\mathtt{Istream}}}
\newcommand{\rstream}{\ensuremath{\mathtt{Rstream}}}
\newcommand{\dstream}{\ensuremath{\mathtt{Dstream}}}

\newcommand{\tumble}{\ensuremath{\mathtt{Tumble}}}
\newcommand{\tumblestart}{\ensuremath{\mathtt{wstart}}}
\newcommand{\tumbleend}{\ensuremath{\mathtt{wend}}}

\newcommand{\hop}{\ensuremath{\mathtt{Hop}}}
\newcommand{\hopstart}{\ensuremath{\mathtt{wstart}}}
\newcommand{\hopend}{\ensuremath{\mathtt{wend}}}

\newcommand{\undo}{\ensuremath{\mathtt{undo}}}
\newcommand{\ptime}{\ensuremath{\mathtt{ptime}}}
\newcommand{\ver}{\ensuremath{\mathtt{ver}}}

\newcommand{\emit}{\ensuremath{\mathtt{EMIT}}}
\newcommand{\emitstream}{\ensuremath{\mathtt{EMIT\,STREAM}}}
\newcommand{\stream}{\ensuremath{\mathtt{STREAM}}}
\newcommand{\afterwatermark}{\ensuremath{\mathtt{AFTER\,WATERMARK}}}
\newcommand{\afterdelay}{\ensuremath{\mathtt{AFTER\,DELAY}}}
\newcommand{\emitafterwatermark}{\ensuremath{\mathtt{EMIT\,AFTER\,WATERMARK}}}
\newcommand{\emitafterdelay}{\ensuremath{\mathtt{EMIT\,AFTER\,DELAY}}}

\lstset{language=SQL,
       showspaces=false,
       basicstyle=\ttfamily\footnotesize,
       numberstyle=\tiny,
       commentstyle=\color{gray},
       captionpos=b,
       morekeywords={EMIT,STREAM,AFTER,WATERMARK,DELAY,MINUTES}}

\section{Introduction}
\label{sec:introduction}

The thesis of this paper, supported by experience developing large open-source frameworks supporting real-world streaming use cases, is that the SQL language and relational model, as-is and with minor non-intrusive extensions, can be very effective for manipulation of streaming data.

Our motivation is two-fold. First, we want to share our observations, innovations, and lessons learned while working on stream processing in widely used open source frameworks. Second, we want to inform the broader database community of the work we are initiating with the international SQL standardization body \cite{incits} to standardize streaming SQL features and extensions, and to facilitate a global dialogue
on this topic (we discuss proposed extensions in Section \ref{sec:standard}). 

\subsection{One SQL for Tables and Streams}
\label{sec:onesql}

Combined, tables and streams cover the critical spectrum of business operations ranging from strategic decision making supported by historical data to near- and real-time data used in interactive analysis. SQL has long been a dominant technology for querying and managing tables of data, backed by decades of research and product development.

We believe, based on our experience and nearly two decades of research on streaming SQL extensions, that using the same SQL semantics in a consistent manner is a productive and elegant way to unify these two modalities of data: it simplifies learning, streamlines adoption, and supports development of cohesive data management systems. Our approach is therefore to present a unified way to manage both tables and streams of data using the same semantics. 

\subsection{Proposed Contribution}
\label{sec:tldr}

Building upon the insights gained in prior art and through our own work, we propose these  contributions in this paper:

    \textbf{Time-varying relations}: First, we propose time-varying relations as a common foundation for SQL, underlying classic point-in-time queries, continuously updated views, and novel streaming queries. A time-varying relation is just what it says: a relation that changes over time, which can also be treated as a function, mapping each point in time to a static relation. Critically, the full suite of existing SQL operators remain valid on time-varying relations (by the natural pointwise application), providing maximal functionality with minimal cognitive overhead. Section \ref{sec:tvr} explores this in detail and Section \ref{sec:standard-tvr} discusses its relationship to standard SQL.

    \textbf{Event time semantics}: Second, we present a concise proposal for enabling robust event time streaming semantics. The extensions we propose preserve all existing SQL semantics and fit in well. By virtue of utilizing time-varying relations as the underlying primitive concept, we can freely combine classical SQL and event time extensions. Section \ref{sec:ets} describes the necessary foundations and Sections \ref{sec:standard-tvr}-\ref{sec:standardwindowing} describe our proposed extensions for supporting event time.

    \textbf{Materialization control}: Third, we propose a modest set of materialization controls to provide the necessary flexibility for handling the breadth of modern streaming use cases.
    
    \begin{itemize}
        \item \textbf{Stream materialization}: To complete the stream-table duality, we propose to allow optionally rendering query output as a stream of changes to the output relation. This stream is, itself, a time-varying relation to which streaming SQL can be applied to express use cases for the stream aspect of systems that differentiate streams and tables. Section \ref{sec:stream-materialization} describes stream materialization in general and Section \ref{sec:standard-stream-materialization} describes our proposal for adding stream materialization to SQL.
    
        \item \textbf{Materialization delay}: Lastly, we propose a minimal, but flexible, framework for delaying the materialization of incomplete, speculative results from a time-varying relation. Our framework adds expressiveness beyond instantaneously updated view semantics to support push semantics (e.g., notification use cases, which are often poorly served by instantaneous update) as well as high-volume use cases (e.g., controlling the frequency of aggregate coalescence in high throughput streams). Section \ref{sec:materialization} discusses materialization control in detail and Section \ref{sec:standardtriggers} presents our framework for materialization.
    \end{itemize}

Taken together, we believe these contributions provide a solid foundation for utilizing the full breadth of standard SQL in a streaming context, while providing additional capabilities for robust event time handling in the context of classic point-in-time queries.

\section{Background and Related Work}
\label{sec:background}

Stream processing and streaming SQL, in their direct forms, as well as under the guise of complex event processing (CEP) \cite{buchmann2009complex} and continuous querying \cite{babu2001continuous}, have been active areas of database research since the 1990s. There have been significant developments in these fields, and we present here a brief survey, by no means exhaustive, of research, industrial, and open source developments relevant to our approach. 

\subsection{A History of Streaming SQL}
\label{sec:history}

Work on stream processing goes back to the introduction of the Tapestry \cite{goldberg1992using} system in 1992, intended for content-based filtering of emails and message board documents using a subset of SQL called TQL  \cite{terry1992tapestry}. Several years later, Liu et al. introduced OpenCQ, an information delivery system driven by user- or application-specified events, the updates of which only occur at specified triggers that don`t require active monitoring or interference \cite{liu1999opencq}. That same group  developed CONQUER, an update/capture system for efficiently monitoring continuous queries over the web using a three-tier architecture designed to share information among variously structured data sources \cite{liu1999conquer}. Shortly thereafter NiagaraCQ emerged, an XML-QL-based query system designed to address scalability issues of continuous queries by grouping similar continuous queries together via dynamic regrouping \cite{chen2000niagara}. OpenCQ, CONQUER, and NiagaraCQ each support arrival and timer-based queries over a large network (i.e. the Internet). However, neither Tapestry nor OpenCQ address multiple query optimization, and NiagaraCQ ignores query execution timings and doesn't specify time intervals \cite{kitagawa2010qe}.

In 2003, Arasu, Babu and Widom introduced the Continuous Query Language (CQL), a declarative language similar to SQL and developed by the STREAM project team at Stanford University. A cohesive syntax, or transformation logic, to process both streaming and static data, this work was the first to introduce an exact semantics for general-purpose, declarative continuous queries over streams and relations. It formalized a form of streams, updateable relations, and their relationships; moreover, it defined abstract semantics for continuous queries constructed on top of relational query language concepts \cite{arasu2003cql,arasu2006cql}. 
\subsubsection{CQL Operators}
CQL defines three classes of operators: relation-to-relation, stream-to-relation, and relation-to-stream. The core operators, relation-to-relation, use a notation similar to SQL. Stream-to-relation operators extract relations from streams using windowing specifications, such as sliding and tumbling windows. Relation-to-stream operators include the $\istream$ (insert stream), $\dstream$ (delete stream), and $\rstream$ (relation stream) operators \cite{arasu2016stream}. Specifically, these three special operators are defined as follows:
\begin{enumerate}
\item $\istream (R)$ contains all $(r, T)$ where $r$ is an element of $R$ at $T$ but not $T-1$
\item $\dstream(R)$ contains all $(r, T)$ where $r$ is an element of $R$ at $T-1$ but not at $T$
\item $\rstream(R)$ contains all $(r, T)$ where r is an element of $R$ at time $T$ \cite{streamprez}
\end{enumerate}
The kernel of many ideas lies within these operators. Notably, time is implicit. The STREAM system accommodates out-of-order data by buffering it on intake and presenting it to the query processor in timestamp order, so the CQL language does not address querying of out-of-order data. 

An important limitation of CQL is that time refers to a logical clock that tracks the evolution of relations and streams, not time as expressed in the data being analyzed, which means time is not a first-class entity one can observe and manipulate alongside other data.

\subsubsection{Other Developments}
The Aurora system was designed around the same time as STREAM to combine archival, spanning, and real-time monitoring applications into one framework. Like STREAM, queries are structured as Directed Acyclic Graphs with operator vertices and data flow edges \cite{abadi2003aurora}. Aurora was used as the query processing engine for Medusa, a load management system for distributed stream processing systems \cite{balazinska2004medusa}, and Borealis, a stream processing engine developed by Brandeis, Brown and MIT. Borealis uses Medusa's load management system and introduced a new means to explore fault-tolerance techniques (results revision and query modification) and dynamic load distribution \cite{abadi2005borealis}. The optimization processes of these systems still do not take event specification into account. Aurora's GUI provides custom operators, designed to handle delayed or missing data, with four specific novel functions: timeout capability, out-of-order input handling, user-defined extendibility, and a resampling operator \cite{cetintemel2003aurora}. These operators are partially based on linear algebra / SQL, but also borrow from AQuery and SEQ \cite{balakrishnan2004aurora}.

IBM introduced SPADE, also known as System S \cite{Gedik2008Spade}, in 2008; this later evolved into InfoSphere Streams, a stream analysis platform which uses SPL, its own native processing language, which allows for event-time annotation.

\subsection{Contemporary Streaming Systems}
\label{sec:contemporary}
\label{sec:legacy}

While streaming SQL has been an area of active research for almost three decades, stream processing itself has enjoyed recent industry attention, and many current streaming systems have adopted some form of SQL functionality.

\textbf{Apache Spark} Spark's Dataset API is a high-level declarative API built on top of Spark SQL's optimizer and execution engine. Dataset programs can be executed on finite data or on streaming data. The streaming variant of the Dataset API is called Structured Streaming \cite{armbrust2018structured}. Structured Streaming queries are incrementally evaluated and by default processed using a micro-batch execution engine, which processes data streams as a series of small batch jobs and features exactly-once fault-tolerance guarantees.

\textbf{KSQL} Confluent's KSQL \cite{confluent2018ksql} is built on top of Kafka Streams, the stream processing framework of the Apache Kafka project. KSQL is a declarative wrapper around Kafka Streams and defines a custom SQL-like syntax to expose the idea of streams and tables \cite{sax2018streams}. KSQL focuses on eventually consistent, materialized view semantics.  

\textbf{Apache Flink} \cite{asf2018flink} features two relational APIs, the LINQ-style \cite{meijer2006linq} Table API and SQL, the latter of which has been adopted by enterprises like Alibaba, Huawei, Lyft, Uber, and others. Queries in both APIs are translated into a common logical plan representation and optimized using Apache Calcite \cite{asf2018calcite}, then optimized and execute as batch or streaming applications.

\textbf{Apache Beam} \cite{asf2018beam} has recently added SQL support, developed with a careful eye towards Beam's unification of bounded and unbounded data processing \cite{akidau2015dataflow}. Beam currently implements a subset of the semantics proposed by this paper, and many of the proposed extensions have been informed by our experiences with Beam over the years.

\textbf{Apache Calcite} \cite{asf2018calcite} is widely used as a streaming SQL parser and planner/optimizer, notably in Flink SQL and Beam SQL. In addition to SQL parsing, planning and optimization, Apache Calcite supports stream processing semantics which have, along with the approaches from Flink and Beam, influenced the work presented in this paper.



There are many other such systems that have added some degree of SQL or SQL-like functionality. A key difference in our new proposal in this work is that other systems are either limited to a subset of standard SQL or bound to specialized operators. Additionally, the other prominent implementations do not fully support robust event time semantics, which is foundational to our proposal. 

In this paper, we synthesize the lessons learned from work on three of these systems - Flink, Beam, Calcite - into a new proposal for extending the SQL standard with the most essential aspects of streaming relational processing.

\section{Minimal streaming SQL foundations}
\label{sec:foundation}

Our proposal for streaming SQL comes in two parts. The first, in this section, is conceptual groundwork, laying out concepts and implementation techniques that support the fundamentals of streaming operations. The second, in Section \ref{sec:standard}, builds on these foundations, identifies the ways in which standard SQL already supports streaming, and proposes minimal extensions to SQL to provide robust support for the remaining concepts. The intervening sections are dedicated to discussing the foundations through examples and lessons learned from our open source frameworks.

\subsection{Time-Varying Relations}
\label{sec:tvr}

In the context of streaming, the key additional dimension to consider is that of time. When dealing with classic relations, one deals with relations \textit{at a single point in time}. When dealing with streaming relations, one must deal with relations as they evolve \textit{over time}. We propose making it explicit that SQL operates over \emph{time-varying relations}, or TVRs.

A time-varying relation is exactly what the name implies: a relation whose contents may vary over time. The idea is compatible with the mutable database tables with which we are already familiar; to a consumer of such a table, it is already a time-varying relation.\footnote{And indeed, the \texttt{AS OF SYSTEM TIME} construct already enshrines the concept of a time-varying relation in the SQL standard.} But such a consumer is explicitly denied the ability to observe or compute based on how the relation changes over time. A traditional SQL query or view can express a derived time-varying relation that evolves in lock step with its inputs: at every point in time, it is equivalent to querying its inputs at exactly that point in time. But there exist TVRs that cannot be expressed in this way, where time itself is a critical input.

TVRs are not a new idea; they are explored in \cite{arasu2003cql,arasu2006cql,sax2018streams}. An important aspect of TVRs is that they may be encoded or materialized in many ways, notably as a sequence of classic relations (\textit{instantaneous relations}, in the CQL parlance), or as a sequence of \texttt{INSERT} and \texttt{DELETE} operations. These two encodings are duals of one another, and correspond to the tables and streams well described by Sax et al. \cite{sax2018streams}. There are other useful encodings based on relation column properties. For example, when an aggregation is invertible, a TVR's encoding may use aggregation differences rather than entire deletes and additions.

Our main contribution regarding TVRs is to suggest that neither the CQL nor the \textit{Streams and Tables} approaches go far enough: rather than defining the duality of streams and tables and then proceeding to treat the two as largely different, we should use that duality to our advantage. The key insight, stated but under-utilized in prior work, is that streams and tables are two representations for one semantic object. This is not to say that the representation itself is not interesting - there are use cases for materializing and operating on the stream of changes itself - but this is again a TVR and can be treated uniformly. 

What's important here is that the core semantic object for relations over time is always the TVR, which by definition supports the entire suite of relational operators, even in scenarios involving streaming data. This is critical, because it means anyone who understands enough SQL to solve a problem in a non-streaming context still has the knowledge required to solve the problem in a streaming context as well.

\subsection{Event Time Semantics}
\label{sec:ets}

Our second contribution deals with event time semantics. Many approaches fall short of dealing with the inherent independence of event time and processing time. The simplest failure is to assume data is ordered according to event time. In the presence of mobile applications, distributed systems, or even just sharded archival data, this is not the case. Even if data is in order according to event time, the progression of a logical clock or processing clock is unrelated to the scale of time as the events actually happened -- one hour of processing time has no relation to one hour of event time. Event time must be explicitly accounted for to achieve correct results.

The STREAM system includes \emph{heartbeats} as an optional feature to buffer out-of-order data and feed it in-order to the query processor. This introduces latency to allow timestamp skew. Millwheel \cite{akidau2013millwheel} based its processing instead on \emph{watermarks}, directly computing on the out-of-order data along with metadata about how complete the input was believed to be. This approach was further extended in Google's Cloud Dataflow \cite{akidau2015dataflow}, which pioneered the out-of-order processing model adopted in both Beam and Flink. 

The approach taken in KSQL \cite{sax2018streams} is also to process the data in arrival order. Its windowing syntax is bound to specific types of event-time windowing\footnote{Not to be confused with SQL's windowing concept.} implementations provided by the system (rather than allowing arbitrary, declarative construction via SQL). Due to its lack of support for watermarks, it is unsuitable for use cases like notifications where some notion of completeness is required, instead favoring an eventual consistency with a polling approach. We believe a more general approach is necessary to serve the full breadth of streaming use cases.

We propose to support event time semantics via two concepts: explicit event timestamps and watermarks. Together, these allow correct event time calculation, such as grouping into intervals (or windows) of event time, to be effectively expressed and carried out without consuming unbounded resources. 

\subsubsection{Event Timestamps}
\label{sec:timestamps}

To perform robust stream processing over a time-varying relation, the rows of the relation should be timestamped in event time and processed accordingly, not in arrival order or processing time.
    
\subsubsection{Watermarks}
\label{sec:watermarks}

A watermark is a mechanism in stream processing for deterministically or heuristically defining a temporal margin of completeness for a timestamped event stream. Such margins are used to reason about the completeness of input data being fed into temporal aggregations, allowing the outputs of such aggregates to be materialized and resources to be released only when the input data for the aggregation are sufficiently complete. For example, a watermark might be compared against the end time of an auction to determine when all valid bids for said auction have arrived, even in a system where events can arrive highly out of order. Some systems provide configuration to allow sufficient slack time for events to arrive.
    
More formally, a watermark is a monotonic function from processing time to event time. For each moment in processing time, the watermark specifies the event timestamp up to which the input is believed to be complete at that point in processing time. In other words, if a watermark observed at processing time $y$ has value of event time $x$, it is an assertion that as of processing time $y$, all future records will have event timestamps greater than $x$. 

\subsection{Materialization Controls}
Our third contribution deals with shaping the way relations are materialized, providing control over \textit{how} the relation is rendered and \textit{when} rows themselves are materialized.

\subsubsection{Stream Materialization}
\label{sec:stream-materialization}

As described in \cite{sax2018streams}, stream changelogs are a space-efficient way of describing the evolution of a TVR over time. Changelogs capture the element-by-element differences between two versions of a relation, in effect encoding the sequence of INSERT and DELETE statements used to mutate the relation over time. They also expose metadata about the evolution of the rows in the relation over time. For example: which rows are added or retracted, the processing time at which a row was materialized, and the revision index of a row for a given event-time interval\footnote{For a more exhaustive look at the types of changelog metadata one might encounter, consult Beam Java's \texttt{PaneInfo} class \cite{paneinfo}.}.

If dealing exclusively in TVRs, as recommended above, rendering a changelog stream of a TVR is primarily needed when materializing a stream-oriented view of that TVR for storage, transmission, or introspection (in particular, for inspecting metadata about the stream such as whether a change was additive or retractive). Unlike other approaches which treat stream changelogs as wholly different objects from relations (and the primary construct for dealing with relations over time), we propose representing the changelog as simply another time-varying relation. In that way, it can be operated on using the same machinery as a normal relation. Furthermore, it remains possible to declaratively convert the changelog stream view back into the original TVR using standard SQL (no special operators needed), while also supporting the materialization delays described next.

\subsubsection{Materialization Delay}
\label{sec:materialization}

By modeling input tables and streams as time-varying relations, and the output of a query as a resulting time-varying relation, it may seem natural to define a query's output as instantaneously changing to reflect any new input. But as an implementation strategy, this is woefully inefficient, producing an enormous volume of irrelevant updates for consumers that are only interested in final results. Even if a consumer is prepared for speculative non-final results, there is likely a maximum frequency that is useful. For example, for a real-time dashboard viewed by a human operator, updates on the order of second are probably sufficient. For top-level queries that are stored or transmitted for external consumption, how frequently and why output materialization occurs is fundamental business logic.

There are undoubtedly many interesting ways to specify when materialization is desired. In Section \ref{sec:standard-materialization} we make a concrete proposal based on experience with real-world use cases. But what is important is that the user has \emph{some} way to express their requirements.

\section{A Motivating Example}
\label{sec:example}

To illustrate the concepts in Section \ref{sec:foundation}, this section examines a concrete example query from the streaming SQL literature. We show how the concepts are used in the query and then walk through its semantics on realistic input.

The following example is from the NEXMark benchmark \cite{tucker2008nexmark} which was designed to measure the performance of stream query systems. The NEXMark benchmark extends the XMark benchmark \cite{schmidt2002xmark} and models an online auction platform where users can start auctions for items and bid on items. The NEXMark data model consists of three streams, \texttt{Person}, \texttt{Auction}, and \texttt{Bid}, and a static \texttt{Category} table that holds details about items.

From the NEXMark benchmark we chose Query 7, defined as: \textit{"Query 7 monitors the highest price items currently on auction. Every ten minutes, this query returns the highest bid (and associated itemid) in the most recent ten minutes."} \cite{tucker2008nexmark}. This is a continuously evaluated query which consumes a stream of bids as input and produces as output a stream of aggregates computed from finite windows of the input. 

Before we show a solution based on plain SQL, we present a variant \cite{nexmarkcql} built with CQL \cite{arasu2003cql} to define the semantics of Query 7:

\begin{lstlisting}[basicstyle=\ttfamily\footnotesize,caption=NEXMark Query 7 in CQL,captionpos=b]
SELECT 
  Rstream(B.price, B.itemid)
FROM 
  Bid [RANGE 10 MINUTE SLIDE 10 MINUTE] B
WHERE 
  B.price = 
    (SELECT MAX(B1.price) FROM BID
     [RANGE 10 MINUTE SLIDE 10 MINUTE] B1);
\end{lstlisting}

Every ten minutes, the query processes the bids of the previous ten minutes. It computes the highest price of the last ten minutes (subquery) and uses the value to select the highest bid of the last ten minutes. The result is appended to a stream. We won't delve into the details of CQL's dialect, but to note some aspects which we will not reproduce in our proposal:

\textbf{CQL makes explicit the concept of streams and relations}, providing operators to convert a stream into a relation (\texttt{RANGE} in our example) and operators to convert a relation into a stream ($\rstream$ in our example). Our approach is based on the single concept of a time-varying relation and does not strictly require conversion operators. 

\textbf{Time is implicit}; the grouping into ten minute windows depends on timestamps that are attached to rows by the underlying stream as metadata. As discussed in Section \ref{sec:ets}, STREAM supports out-of-order timestamps by buffering and feeding to CQL in order so intervals of event time always correspond to contiguous sections of the stream. Our approach is to process out-of-order data directly by making event timestamps explicit and leveraging watermarks to reason about input completeness.

\textbf{Time moves in lock step for the whole query.} There is no explicit condition that the window in the subquery corresponds to the window in the main query. We make this relationship explicit via a join condition.

In contrast, here is Query 7 specified with our proposed extensions to standard SQL.\footnote{Note that the query as written here is an evolution beyond what is currently supported in Flink and Beam via Calcite; we discuss the differences in Section \ref{sec:standard} and Appendix \ref{appendix:frameworks}.}

\begin{lstlisting}[
    basicstyle=\ttfamily\footnotesize,
    caption=NEXMark Query 7 in SQL,
    label=lst:nexmarkSql]
SELECT
  MaxBid.wstart, MaxBid.wend,
  Bid.bidtime, Bid.price, Bid.itemid
FROM 
  Bid,
  (SELECT
     MAX(TumbleBid.price) maxPrice, 
     TumbleBid.wstart wstart,
     TumbleBid.wend wend
   FROM
     Tumble(
       data    => TABLE(Bid),
       timecol => DESCRIPTOR(bidtime)
       dur     => INTERVAL '10' MINUTE) TumbleBid
   GROUP BY
     TumbleBid.wend) MaxBid
WHERE 
  Bid.price = MaxBid.maxPrice AND
  Bid.bidtime >= MaxBid.wend 
                 - INTERVAL '10' MINUTE AND
  Bid.bidtime < MaxBid.wend;
\end{lstlisting}

This query computes the same result, but does so using our proposed extensions to standard SQL (as well as SQL standard features from 2016). Noteworthy points:

\textbf{The column \texttt{bidtime} holds the time at which a bid occurred.} In contrast to the prior query, timestamps are explicit data. Rows in the \texttt{Bid} stream do \emph{not} arrive in order of \texttt{bidtime}.

\textbf{The \texttt{Bid} stream is presumed to have a watermark}, as described in Section \ref{sec:ets}, estimating completeness of \texttt{BidTime} as a lower bound on future timestamps in the \texttt{bidtime} column. Note that the requirement does not affect the basic semantics of the query. The same query can be evaluated without watermarks over a table that was recorded from the \texttt{bid} stream, yielding the same result.

\textbf{$\tumble$ is a table-valued function} \cite{SQL2016} \textbf{which assigns each row in the bid stream to the 10-minute interval containing \texttt{bidtime}.} The output table \texttt{TumbleBid} has all the same columns as \texttt{Bid} plus two additional columns $\tumblestart$ and $\tumbleend$, which repesent the start and end of the tumbling window interval, respectively. The $\tumbleend$ column contains timestamps and has an associated watermark that estimates completeness of \texttt{TumbleBid} relative to $\tumbleend$.
    
\textbf{The \texttt{GROUP BY TumbleBid.wend} clause is where the watermark is used.} Because the watermark provides a lower bound on not-yet-seen values for $\tumbleend$, it allows an implementation to reason about when a particular grouping of inputs is complete. This fact can be used to delay materialization of results until aggregates are known complete, or to provide metadata indicating as much.

\textbf{As the \texttt{Bid} relation evolves over time, with new events being added, the relation defined by this query also evolves.} This is identical to instantaneous view semantics. We have not used the advanced feature of managing the materialization of this query.

Now let us apply this query to a concrete dataset to illustrate how it might be executed. As we're interested in streaming data, we care not only about the data involved, but also \textit{when} the system becomes aware of them (processing time), as well as \textit{where} in event time they occurred, and the system's own understanding of input completeness in the event-time domain (i.e., the watermark) over time. The example data set we will use is the following:

\footnotesize
\begin{verbatim}
    8:07    WM -> 8:05
    8:08    INSERT (8:07, $2, A)
    8:12    INSERT (8:11, $3, B)
    8:13    INSERT (8:05, $4, C)
    8:14    WM -> 8:08
    8:15    INSERT (8:09, $5, D)
    8:16    WM -> 8:12
    8:17    INSERT (8:13, $1, E)
    8:18    INSERT (8:17, $6, F)
    8:21    WM -> 8:20
\end{verbatim}
\normalsize

Here, the left column of times includes the processing times at which events occur within the system. The right column describes the events themselves, which are either the watermark advancing to a point in event time or a \texttt{(bidtime, price, item)} tuple being inserted into the stream.

The example SQL query in Listing \ref{lst:nexmarkSql} would yield the following results when executed on this dataset at time 8:21 (eliding most of the query body for brevity):

\begin{lstlisting}[
    caption=NEXMark Query 7 over full dataset,
    label=lst:nexmark-default]
8:21> SELECT ...;
------------------------------------------
| wstart | wend | bidtime | price | item |
------------------------------------------
| 8:00   | 8:10 | 8:09    |    $5 |    D |
| 8:10   | 8:20 | 8:17    |    $6 |    F |
------------------------------------------
\end{lstlisting}

This is effectively the same output that would have been provided by the original CQL query, with the addition of explicit window start, window end, and event occurrence timestamps. 

However, this is a table view of the data set capturing a point-in-time view of the entire relation at query time, not a stream view. If we were to have executed this query earlier in processing time, say at 8:13, it would have looked very different due to the fact that only half of the input data had arrived by that time:
\begin{lstlisting}[
  caption=NEXMark Query 7 over partial dataset,
  label=lst:nexmark-earlier]
8:13> SELECT ...;
------------------------------------------
| wstart | wend | bidtime | price | item |
------------------------------------------
| 8:00   | 8:10 | 8:05    |    $4 |    C |
| 8:10   | 8:20 | 8:11    |    $3 |    B |
------------------------------------------
\end{lstlisting}

In Section 6, we'll describe how to write a query that creates a stream of output matching that from the original CQL query, and also why our approach is more flexible overall.

\section{Lessons Learned in Practice}

Our proposed SQL extensions are informed by prior art and related work, and derived from experience in working on Apache Calcite, Flink, and Beam -- open source frameworks with wide adoption across the industry and by other open source frameworks.

In Appendix \ref{appendix:frameworks}, we describe the general architectural properties of these three frameworks, the breadth of their adoption, and streaming implementations that exist today. Though the implementations thus far fall short of the full proposal we make in Section \ref{sec:standard}, they are a step in the right direction and have yielded useful lessons which have informed its evolution. Here we summarize those lessons:

\textbf{Some operations only work (efficiently) on watermarked event time attributes.} Whether performing an aggregation on behalf of the user or executing overtly stateful business logic, an implementer must have a way to maintain finite state over infinite input. Event time semantics, particularly watermarks, are critical. State for an ongoing aggregation or stateful operator can be freed when the watermark is sufficiently advanced that the state won't be accessed again.

\textbf{Operators may erase watermark alignment of event time attributes.} Event time processing requires that event timestamps are aligned with watermarks. Since event timestamps are exposed as regular attributes, they can be referenced in arbitrary expressions. Depending on the expression, the result may or may not remain aligned with the watermarks; 
these cases need to be taken into account during query planning. In some cases it is possible to preserve watermark alignment by adjusting the watermarks, and in others an event time attribute loses its special property.

\textbf{Time-varying relations might have more than one event time attribute.} Most stream processing systems that feature event time processing only support a single event time attribute with watermarks. When joining two TVRs it can happen that the event time attributes of both input TVRs are preserved in the resulting TVR. One approach to address this situation is to "hold-back" the watermark such that all event time attributes remain aligned.

\textbf{Reasoning about what can be done with an event time attribute can be difficult for users.} In order to define a query that can be efficiently executed using event time semantics and reasoning, event time attributes need to be used at specific positions in certain clauses, for instance as an \texttt{ORDER BY} attribute in an \texttt{OVER} clause. These positions are not always easy to spot and failing to use event time attributes correctly  easily leads to very expensive execution plans with undesirable semantics.

\textbf{Reasoning about the size of query state is sometimes a necessary evil.} Ideally, users should not need to worry about internals when using SQL. However, when consuming unbounded input user intervention is useful or sometimes necessary. So we need to consider what metadata the user needs to provide (active interval for attribute inserts or updates, e.g. sessionId) and also how to give the user feedback about the state being consumed, relating the physical computation back to their query.

\textbf{It is useful for users to distinguish between streaming and materializing operators}. In Flink and Beam, users need to reason explicitly about which operators may produce updating results, which operators can consume updating results, and the effect of operators on event time attributes. These low-level considerations are inappropriate for SQL and have no natural place in relational semantics; we need materialization control extensions that work well with SQL.

\textbf{Torrents of updates}: For a high-throughput stream, it is very expensive to issue updates continually for all derived values. Through materialization controls in Flink and Beam, this can be limited to fewer and more relevant updates.

\section{Extending the SQL Standard}
\label{sec:standard}
\label{sec:standardplease}

Work presented here is part of an initial effort to standardize streaming SQL and define our emerging position on its features.
In this section, we will first briefly discuss some ways in which SQL already supports streaming, after which we will present our proposed streaming extensions.

\subsection{Existing Support for Streaming in SQL}

SQL as it exists today already includes support for a number of streaming related approaches. Though not sufficient to cover all relevant streaming use cases, they provide a good foundation upon which to build, notably:

\textbf{Queries are on table snapshots}:
    As a classical SQL table evolves, queries can execute on their current contents. In this way, SQL already plays nicely with relations over time, albeit only in the context of static snapshots.

\textbf{Materialized Views}:
    Views (semantically) and materialized views (physically) map a query pointwise over a TVR. At any moment, the view is the result of a query applied to its inputs at that moment. This is an extremely useful initial step in stream processing.

\textbf{Temporal tables}:
    Temporal tables embody the idea of a time-varying relation, and provide the ability to query snapshots of the table from arbitrary points of time in the past via \texttt{AS OF SYSTEM TIME operators}.

\textbf{MATCH RECOGNIZE}:
    The \texttt{MATCH\_RECOGNIZE} clause was added with SQL:2016 \cite{SQL2016}. When combined with event time semantics, this extension is highly relevant to streaming SQL as it enables a new class of stream processing use case, namely complex event processing and pattern matching \cite{buchmann2009complex}.

\subsection{Time-Varying Relations, Event Time Columns, and Watermarks}
\label{sec:standard-tvr}

There are no extensions necessary to support time-varying relations. Relational operators as they exist today already map one time-varying relation to another naturally. 

To enable event time semantics in SQL, a relation may include in its schema columns that contain event timestamps. Query execution requires knowledge of which column(s) correspond to event timestamps to associate them with watermarks, described below. The metadata that a column contains event timestamps is to be stored as part of or alongside the schema. The timestamps themselves are used in a query like any other data, in contrast to CQL where timestamps themselves are metadata and KSQL which implicitly references event time attributes that are declared with the schema.

To support unbounded use cases, watermarks are also available as semantic inputs to standard SQL operators. This expands the universe of relational operators to include operators that are not pointwise with respect to time, as in Section \ref{sec:standardtriggers}. For example, rows may be added to an output relation based only on the advancement of the watermark, even when no rows have changed in the input relation(s).

\begin{extension}[Watermarked event time column]
An event time column in a relation is a distinguished column of type TIMESTAMP with an associated watermark. The watermark associated with an event time column is maintained by the system as time-varying metadata for the relation as a whole and provides a lower bound on event timestamps that may be added to the column.
\end{extension}

\subsection{Grouping on Event Timestamps}

When processing over an unbounded stream, an aggregate projected in a query of the form \texttt{SELECT ... GROUP BY ...} is complete when it is known that no more rows will contribute to the aggregate. Without extensions, it is never known whether there may be more inputs that contribute to a grouping. Under event time semantics, the watermark gives a measure of completeness and can determine when a grouping is complete based on event time columns. This corresponds to the now-widespread notion of \emph{event-time windowing}. We can adapt this to SQL by leveraging event time columns and watermarks.

\begin{extension}[Grouping on event timestamps]
When a GROUP BY clause contains a grouping key that is an event time column, any grouping where the key is less than the watermark for that column is declared complete, and further inputs that would contribute to that group are dropped (in practice, a configurable amount of allowed lateness is often needed, but such a mechanism is beyond the scope of this paper; for more details see Chapter 2 of \cite{akidau2018streaming})  Every \texttt{GROUP BY} clause with an unbounded input is required to include at least one event-time column as a grouping key.
\end{extension}

\subsection{Event-Time Windowing Functions}
\label{sec:standardwindowing}

It is rare to group by an event time column that contains original event timestamps unless you are trying to find simultaneous events. Instead, event timestamps are usually mapped to a distinguished end time after which the grouping is completed. In the example from Section \ref{sec:example}, bid timestamps are mapped to the end of the ten minute interval that contains them. We propose adding built-in table-valued functions that augment a relation with additional event timestamp columns for these common use cases (while leaving the door open for additional built-in or custom TVFs in the future).

\begin{extension}[Event-time windowing functions]
Add (as a starting point) built-in table-valued functions \texttt{Tumble} and \texttt{Hop} which take a relation and event time column descriptor as input and return a relation with additional event-time interval columns as output, and establish a convention for the event-time interval column names.
\end{extension}

The invocation and semantics of \texttt{Tumble} and \texttt{Hop} are below. There are other useful event time windowing functions used in streaming applications which the SQL standard may consider adding, but these two are extremely common and illustrative. For brevity, we show abbreviated function signatures and describe the parameters in prose, then illustrate with example invocations.

\subsubsection{Tumble}

Tumbling (or "fixed") windows partition event time
into equally spaced disjoint covering intervals. $\tumble$ takes three required parameters and one optional parameter:

\begin{lstlisting}[
       language=SQL,
       showspaces=false,
       basicstyle=\ttfamily\small,
       numberstyle=\tiny,
       commentstyle=\color{gray}
    ]
Tumble(data, timecol, dur, [offset])
\end{lstlisting}

\begin{itemize}
    \item \texttt{data} is a table parameter that can be any relation with an event time column.
    \item \texttt{timecol} is a column descriptor indicating which event time column of \texttt{data} should be mapped to tumbling windows.
    \item \texttt{dur} is a duration specifying the width of the tumbling windows.
    \item \texttt{offset} (optional) specifies that the tumbling should begin from an instant other than the standard beginning of the epoch.
\end{itemize}

The return value of $\tumble$ is a relation that includes all columns of \texttt{data} as well as additional event time columns $\tumblestart$ and $\tumbleend$. Here is an example invocation on the \texttt{Bid} table from the example in Section \ref{sec:example}:

\begin{lstlisting}[caption=Applying the Tumble TVF]
8:21> SELECT * 
      FROM Tumble(
        data    => TABLE(Bid),
        timecol => DESCRIPTOR(bidtime),
        dur     => INTERVAL '10' MINUTES,
        offset  => INTERVAL '0' MINUTES);
------------------------------------------
| wstart | wend | bidtime | price | item |
------------------------------------------
| 8:00   | 8:10 | 8:07    |    $2 |    A |
| 8:10   | 8:20 | 8:11    |    $3 |    B |
| 8:00   | 8:10 | 8:05    |    $4 |    C |
| 8:00   | 8:10 | 8:09    |    $5 |    D |
| 8:10   | 8:20 | 8:13    |    $1 |    E |
| 8:10   | 8:20 | 8:17    |    $6 |    F |
------------------------------------------
\end{lstlisting}

Users can group by $\tumblestart$ or $\tumbleend$; both result in the same groupings and, assuming ideal watermark propagation, the groupings reach completeness at the same time. For example, grouping by $\tumbleend$:

\begin{lstlisting}[caption=Tumble combined with GROUP BY]
8:21> SELECT MAX(wstart), wend, SUM(price) 
      FROM Tumble(
        data    => TABLE(Bid),
        timecol => DESCRIPTOR(bidtime),
        dur     => INTERVAL '10' MINUTES)
      GROUP BY wend;
-------------------------
| wstart | wend | price |
-------------------------
| 8:00   | 8:10 |   $11 |
| 8:10   | 8:20 |   $10 |
-------------------------
\end{lstlisting}

\subsubsection{Hop}
Hopping (or "sliding") event time windows place intervals of a fixed size evenly spaced across event time. $\hop$ takes four required parameters and one optional parameter. All parameters are analogous to those for $\tumble$ except for \texttt{hopsize}, which specifies the duration between the starting points (and endpoints) of the hopping windows, allowing for overlapping windows ($hopsize < dur$, common) or gaps in the data ($hopsize > dur$, rarely useful).

The return value of $\hop$ is a relation that includes all columns of \texttt{data} as well as additional event time columns $\hopstart$ and $\hopend$. Here is an example invocation on the \texttt{Bid} table from the example in Section \ref{sec:example}:

\begin{lstlisting}[caption=Applying the Hop TVF]
8:21> SELECT * 
      FROM Hop(
        data    => TABLE Bids,
        timecol => DESCRIPTOR(bidtime),
        dur     => INTERVAL '10' MINUTES,
        hopsize => INTERVAL '5' MINUTES);
------------------------------------------
| wstart | wend | bidtime | price | item |
------------------------------------------
| 8:00   | 8:10 | 8:07    |    $2 |    A |
| 8:05   | 8:15 | 8:07    |    $2 |    A |
| 8:05   | 8:15 | 8:11    |    $3 |    B |
| 8:10   | 8:20 | 8:11    |    $3 |    B |
| 8:00   | 8:10 | 8:05    |    $4 |    C |
| 8:05   | 8:15 | 8:05    |    $4 |    C |
| 8:00   | 8:10 | 8:09    |    $5 |    D |
| 8:05   | 8:15 | 8:09    |    $5 |    D |
| 8:05   | 8:15 | 8:13    |    $1 |    E |
| 8:10   | 8:20 | 8:13    |    $1 |    E |
| 8:10   | 8:20 | 8:17    |    $6 |    F |
| 8:15   | 8:25 | 8:17    |    $6 |    F |
------------------------------------------
\end{lstlisting}
    
Users can group by $\hopstart$ or $\hopend$ with the same effect, as with tumbling windows. For example:

\begin{lstlisting}[caption=Hop combined with GROUP BY]
8:21> SELECT MAX(wstart), wend, SUM(price) 
      FROM Hop(
        data    => TABLE (Bid),
        timecol => DESCRIPTOR(bidtime),
        dur     => INTERVAL '10' MINUTES,
        hopsize => INTERVAL '5'  MINUTES)
      GROUP BY wend;
-------------------------
| wstart | wend | price |
-------------------------
| 8:00   | 8:10 |   $11 |
| 8:05   | 8:15 |   $15 |
| 8:10   | 8:20 |   $10 |
| 8:15   | 8:25 |    $6 |
-------------------------
\end{lstlisting}

Using table-valued functions improves on the current state of implementations in the following ways:

\textbf{\texttt{GROUP BY} is truly a grouping of rows according to a column's value.} In Calcite, Beam, and Flink, \texttt{GROUP BY HOP(...)} violates relational semantics by causing multiple input rows.
    
\textbf{A more uniform notation for all window functions.} The near-trivial $\tumble$ has the same general form as the input-expanding $\hop$, and using a table-valued functions allows adding a wide variety of more complex functionality (such as calendar windows or sessionization) with a similar look-and-feel.
    
\textbf{Engines have flexibility in how they implement these table-valued functions.} Rows in the output may appear and disappear as appropriate according to downstream materialization requirements.

\subsection{Materialization Controls}
\label{sec:standard-materialization}

The last piece of our proposal centers around materialization controls, allowing users flexibility in shaping \textit{how} and \textit{when} the rows in their TVRs are materialized over time.

\subsubsection{Stream Materialization}
\label{sec:standard-stream-materialization}

The \textit{how} aspect of materialization centers around the choice of materializing a TVR as a table or stream. The long-standing default for relations has been to materialize them as tables. And since this approach is completely compatible with the idea of swapping point-in-time relations with time-varying relations, no changes around materializing a table are necessary. However, in certain situations, materializing a stream-oriented changelog view of the TVR is desirable. 

In these cases, we require some way to signal to the system that the changelog of the relation should be materialized. We propose the use of a new $\emitstream$ modifier on a query to do this. Recall our original motivating query results from Listing \ref{lst:nexmark-default}, which rendered a table view of our example query.
By adding $\emitstream$ at the top-level, we materialize the \textit{stream changelog} for the TVR instead of a point-in-time snapshot of the relation itself:

\begin{lstlisting}[
       basicstyle=\ttfamily\tiny,
       caption=Stream changelog materialization
    ]
8:08> SELECT ... EMIT STREAM;
---------------------------------------------------------------
| wstart | wend | bidtime | price | item | undo | ptime | ver |
---------------------------------------------------------------
| 8:00   | 8:10 | 8:07    |    $2 |    A |      | 8:08  |   0 |
| 8:10   | 8:20 | 8:11    |    $3 |    B |      | 8:12  |   0 |
| 8:00   | 8:10 | 8:07    |    $2 |    A | undo | 8:13  |   1 |
| 8:00   | 8:10 | 8:05    |    $4 |    C |      | 8:13  |   2 |
| 8:00   | 8:10 | 8:05    |    $4 |    C | undo | 8:15  |   3 |
| 8:00   | 8:10 | 8:09    |    $5 |    D |      | 8:15  |   4 |
| 8:10   | 8:20 | 8:11    |    $3 |    B | undo | 8:18  |   1 |
| 8:10   | 8:20 | 8:17    |    $6 |    F |      | 8:18  |   2 |
...
\end{lstlisting}

Note that there are a number of additional columns included in the $\stream$ version:

\begin{itemize}
    \item $\undo$: whether the row is a retraction of a previous row or not.
    \item $\ptime$: the processing time offset of the row in the changelog.
    \item $\ver$: a sequence number that versions this row with respect to other rows corresponding to different revisions of the same event-time window.
\end{itemize}

A changelog only has multiple revisions to a row when there is a aggregation present in the query resulting in changes to the row over time.

\begin{extension}[Stream Materialization]
\texttt{EMIT STREAM} results in a time-varying relation representing \emph{changes} to the classical result of the query. In addition to the schema of the classical result, the change stream includes columns indicating: whether or not the row is a retraction of a previous row, the changelog processing time offset of the row, a sequence number relative to other changes to the same event time grouping.
\end{extension}

One could imagine other options, such as allowing materialization of deltas rather than aggregates, or even entire relations a la CQL's $\rstream$. These could be specified with additional modifiers, but are beyond the scope of this paper.

As far as equaling the output of the original CQL query, the $\stream$ keyword is a step in the right direction, but it's clearly more verbose, capturing the full evolution of the highest bidders for the given 10-minute event time windows as data arrive, whereas the CQL version provided only a single answer per 10-minute window once the input data for that window was complete. To tune the stream output to match the behavior of CQL (but accommodating out-of-order input data), we need to support materialization delay.

\subsubsection{Materialization Delay}
\label{sec:standardtriggers}

The \textit{when} aspect of materialization centers around the way relations evolve over time. The standard approach is on a record-by-record basis: as DML operations such as INSERT and DELETE are applied to a relation, those changes are immediately reflected. However, when dealing with aggregate changes in relations, it's often beneficial to delay the materialization of an aggregate in some way. Over the years, we've observed two main categories of delayed materialization in common use: completeness delays and periodic delays.
\\ 

\textit{Completeness delays}: Event time windowing provides a means for slicing an unbounded relation into finite temporal chunks, and for use cases where eventual consistency of windowed aggregates is sufficient, no further extensions are required. However, some use cases dictate that aggregates only be materialized when their inputs are complete, such as queries for which partial results are too unstable to be of any use, such a query which determines if a numeric sum is even or odd.  These still benefit from watermark-driven materialization even when consumed as a table.

Recall again our query from Listing \ref{lst:nexmark-earlier} where we queried the table version of our relation at 8:13. That query presented a partial result for each window, capturing the highest priced items for each tumbling window at that point in processing time. For use cases where presenting such partial results is undesirable, we propose the syntax $\emit\,\afterwatermark$ to ensure the table view would only materialize rows whose input data were complete. In that way, our query at 8:13 would return an empty table:

\begin{lstlisting}[caption=Watermark materialization: incomplete]
8:13> SELECT ... EMIT AFTER WATERMARK;
------------------------------------------
| wstart | wend | bidtime | price | item |
------------------------------------------
------------------------------------------
\end{lstlisting}

If we were to query again at 8:16, once the watermark had passed the end of the first window, we'd see the final result for the first window, but still none for the second:

\begin{lstlisting}[caption=Watermark materialization: partial]
8:16> SELECT ... EMIT AFTER WATERMARK;
------------------------------------------
| wstart | wend | bidtime | price | item |
------------------------------------------
| 8:00   | 8:10 | 8:09    |    $5 |    D |
------------------------------------------
\end{lstlisting}

And then if we queried again at 8:21, after the watermark had passed the end of the second window, we would finally have the final answers for both windows:

\begin{lstlisting}[caption=Watermark materialization: complete]
8:21> SELECT ... EMIT AFTER WATERMARK;
------------------------------------------
| wstart | wend | bidtime | price | item |
------------------------------------------
| 8:00   | 8:10 | 8:09    |    $5 |    D |
| 8:10   | 8:20 | 8:17    |    $6 |    F |
------------------------------------------
\end{lstlisting}

We also can use $\stream$ materialization to concisely observe the evolution of the result, which is analogous to what the original CQL query would produce:
\begin{lstlisting}[caption=Watermark materialization of a stream,basicstyle=\ttfamily\tiny]
8:08> SELECT ... EMIT STREAM AFTER WATERMARK;
---------------------------------------------------------------
| wstart | wend | bidtime | price | item | undo | ptime | ver |
---------------------------------------------------------------
| 8:00   | 8:10 | 8:09    |    $5 |    D |      | 8:16  |   0 |
| 8:10   | 8:20 | 8:17    |    $6 |    F |      | 8:21  |   0 |
...
\end{lstlisting}

Comparing this to the evolution of the streamed changelog in Section \ref{sec:standard-stream-materialization} illustrates the difference with $\afterwatermark$:

\begin{itemize}
    \item There is exactly one row per window, each containing the final result.
    \item The $\ptime$ values no longer correspond to the arrival time of the max bid records, but instead the processing time at which the watermark advanced beyond the end of the given window.
\end{itemize}

The most common example of delayed stream materialization is notification use cases, where polling the contents of an eventually consistent relation is infeasible. In this case, it's more useful to consume the relation as a stream which contains only aggregates whose input data is known to be complete. This is the type of use case targeted by the original CQL top bids query.

\begin{extension}[Materialization Delay: Completeness]
When a query has an $\emitafterwatermark$ modifier, only complete rows from the results are materialized.
\end{extension}

\textit{Periodic delays}: The second delayed materialization use case we care about revolves around managing the verbosity of an eventually consistent $\stream$ changelog. The default $\stream$ rendering, as we saw above, provides updates every time any row in the relation changes. For high volume streams, such a changelog can be quite verbose. In those cases, it is often desirable to limit how frequently aggregates in the relation are updated. To do so, we propose the addition of an $\afterdelay$ modifier to the $\emit$ clause, which dictates a delay imposed on materialization after a change to a given aggregate occurs, for example:

\begin{lstlisting}[
    caption=Periodic delayed stream materialization,
    basicstyle=\ttfamily\tiny
    ]
8:08> SELECT ... EMIT STREAM AFTER DELAY INTERVAL '6' MINUTES;
---------------------------------------------------------------
| wstart | wend | bidtime | price | item | undo | ptime | ver |
---------------------------------------------------------------
| 8:00   | 8:10 | 8:05    |    $4 |    C |      | 8:14  |   0 |
| 8:10   | 8:20 | 8:17    |    $6 |    F |      | 8:18  |   0 |
| 8:00   | 8:10 | 8:05    |    $4 |    C | undo | 8:21  |   1 |
| 8:00   | 8:10 | 8:09    |    $5 |    D |      | 8:21  |   2 |
...
\end{lstlisting}
\normalsize

In this example, multiple updates for each of the windows are compressed together, each within a six-minute delay from the first change to the row.

\begin{extension}[Periodic Materialization]
When a query has $\emitafterdelay\,d$, rows are materialized with period $d$ (instead of continuously).
\end{extension}

It's also possible to combine $\afterdelay$ modifiers with $\afterwatermark$ modifiers to provide the early/on-time/late pattern \cite{akidau2018streaming} of repeated periodic updates for partial result rows, followed by a single on-time row, followed by repeated periodic updates for any late rows.

\begin{extension}[Combined Materialization Delay]
When a query has $\emit\,\afterdelay\,d\,\mathtt{AND}\,\afterwatermark$, rows are materialized with period $d$ as well as when complete.
\end{extension}

\section{Summary}

Streaming SQL is an exercise in manipulating relations over time. The large body of streaming SQL literature combined with recent efforts in the modern streaming community form a strong foundation for basic streaming semantics, but room for improvement remains in the dimensions of usability, flexibility, and robust event-time processing. We believe that the three contributions proposed in this paper, (1) pervasive use of time-varying relations, (2) robust event-time semantics support, and (3) materialization control can substantially improve the ease-of-use of streaming SQL. Moreover, they will broaden the menu of available operators to not only include the full suite of point-in-time relational operators available in standard SQL today, but also extend the capabilities of the language to operators that function over time to shape \textit{when} and \textit{how} relations evolve. 

\section{Future Work}
\textbf{Expanded/custom event-time windowing}: Although the windowing TVFs proposed in Section \ref{sec:standardwindowing} are common, Beam and Flink both provide many more, e.g., transitive closure sessions (periods of contiguous activity), keyed sessions (periods with a common session identifier, with timeout), and calendar-based windows. Experience has also shown that pre-built solutions are never sufficient for all use cases (Chapter 4 of \cite{akidau2018streaming}); ultimately, users should be able to utilize the power of SQL to describe their own custom-windowing TVFs.

\textbf{Time-progressing expressions}: Computing a view over the tail of a stream is common, for example counting the bids of the last hour. Conceptually, this can be done with a predicate like \texttt{(bidtime > CURRENT\_TIME - INTERVAL '1' HOUR)}. However, the SQL standard defines that expressions like \texttt{CURRENT\_TIME} are fixed at query execution time. Hence, we need expressions that progress over time.

\textbf{Correlated access to temporal tables} A common use case in streaming SQL is to enrich a table with attributes from a temporal table at a specific point in time, such as enriching an order with the currency exchange rate at the time when the order was placed. Currently, only a temporal version specified by a fixed literal \texttt{AS OF SYSTEM TIME} can be accessed. To enable temporal tables for joins, the table version needs to be accessible via a correlated join attribute.

\textbf{Streaming changelog options}: As alluded to in Section \ref{sec:standard-stream-materialization}, more options for stream materialization exist, and $\emit$ should probably be extended to support them. In particular, rendering a stream changelog as a sequence of deltas.

\textbf{Nested \texttt{EMIT}}: Though we propose limiting the application of $\emit$ to the top level of a query, an argument can be made for the utility of allowing $\emit$ at any level of nested query. It is worthwhile to explore the tension between additional power and additional complexity this change would impose.

\textbf{Graceful evolution}: Streaming queries by definition exist over an extended period of time, but software is never done nor perfect: bugs are uncovered, requirements evolve, and over time, long-running queries must change. The stateful nature of these queries imposes new challenges regarding the evolution of intermediate state. This remains an unsolved problem for the streaming community in general, but its relevance in the more abstract realm of SQL is all the greater.

\textbf{More rigorous formal definitions of semantics}: Although we've tried to provide semi-formal analyses of concepts presented in this paper where applicable, we as a streaming community still lack  a true formal analysis of what streaming means, particularly when applied to some of the more subtle aspects of event-time processing such as watermarks and materialization controls. A more rigorous survey of modern streaming concepts would be a welcome and beneficial addition to the literature.

\bibliographystyle{ACM-Reference-Format}
\bibliography{bibliography}

\appendix

\section{Acknowledgements}

We would like to thank everyone who has contributed to the Calcite, Flink, and Beam streaming SQL implementations described in this paper, in particular: Robert Bradshaw, Jes\'us Camacho-Rodr\'iguez, Ben Chambers, Hequn Cheng, Xingcan Cui, Lukasz Cwik, Stephan Ewen, Kai Jiang, Xiaowei Jiang, Anton Kedin, Reuven Lax, Xu Mingmin, Michael Mior, Andrew Pilloud, Jincheng Sun, Timo Walther, Rui Wang, Shaoxuan Wang, and James Xu.

We would also like to thank Jeff Shute and Mosha Pasumansky for their help refining our approach, in particular recommending the usage of TVFs.

This manuscript has been in part co-authored by UT-Battelle, LLC under Contract No. DE-AC05-00OR22725 with the U.S. Department of Energy.

\section{Streaming Implementation in Apache Beam, Calcite, and Flink}
\label{appendix:frameworks}

This appendix presents the architectural and implementation details of three open source frameworks with support for unifying SQL: Apache Calcite, the foundation on which SQL support in Flink and Beam are built, and Apache Flink and Beam individually.

Many of the approaches that we presented in this paper are implemented by Apache Beam, Calcite, and Flink. All three projects follow the model of time-varying relations to provide unified semantics for processing static tables and unbounded streams. Apache Beam's and Apache Flink's implementations of SQL leverage event time semantics as a robust framework to handle out-of-order data and to reason about result completeness.

Our proposal for streaming SQL has been adopted by enterprises like Alibaba, Huawei, Lyft, Uber, and others. These companies provide SQL on data streams as public pay-per-use services or as internal services for engineers and analysts. 
From user feedback we identified the following reasons why our approach of streaming SQL was adopted:

\begin{enumerate}
    \item Development and adoption costs are significantly lower compared to non-declarative stream processing APIs.
    \item Familiarity with standard SQL eases adoption compared to non-standardized query languages.
    \item Common stream processing tasks such as windowed aggregations and joins can be easily expressed and efficiently executed due to event time semantics.
    \item In case of faulty application logic or service outages, a recorded data stream can be reprocessed by the same query that processes the live data stream.
\end{enumerate}

We believe that the manifold adoption and deployment in production environments serves as a testimonial for our approach.

\subsection{Streaming SQL in Calcite}
\label{sec:calcite-impl}

Apache Calcite \cite{begoli2018calcite} is an open source query processor that provides common functionality required by database management systems (query processing, optimization, and query language support), except for data storage management. This is intentional, to allow for Calcite to be a mediation engine between applications having one or more data storage locations or multiple data processing engines, as well as the option to build custom data processing systems.

\subsubsection{Parsing, Optimization, and Query Processing}

Calcite's optimizer uses a tree of relational operators as its internal representation. The optimization engine, fundamentally, consists of rules, metadata providers, and planner engines. Calcite optimizes queries by repeatedly applying planner rules to a relational expression. A cost model guides the process, and the planner engine tries to generate an alternative expression that has the same semantics as the original, but at a lower cost. Information is supplied to the optimizer via default (optionally customizable) metadata providers, both guiding the planner toward reducing overall query plan cost and providing information to the optimizer rules as they are applied. 
Calcite contains a query parser and validator that can translate an SQL query to a tree of relational operators; absent a storage layer, Calcite defines table schemas and views in external storage engines via adapters, an architectural pattern that defines how diverse data sources are incorporated.  

\subsubsection{``A Living Lab'' for SQL-based Streaming Semantics}

Calcite supports streaming queries based on a set of CQL-inspired, streaming-specific extensions beyond standard SQL, namely STREAM extensions, windowing extensions, implicit references to streams via window expressions in joins, and others. The STREAM directive focuses on incoming records; in its absence, the query focuses on existing records. Calcite also uses SQL analytic functions to express sliding and cascading window aggregations. 

In order to support streaming queries, Calcite has extended SQL and relational algebra while maintaining the syntactic likeness and idioms of SQL -- we designed Calcite's SQL as an extension to standard SQL, not another SQL-like language. 
This distinction is important for the several reasons, namely i. lowering of learning and adoption barriers -- streaming SQL is easy to learn for anyone who knows regular SQL; ii. the semantics of Streaming SQL are clear -- we aim to produce the same results on a stream as if the same data were in a table; iii. users can write queries that combine streams and tables, or the history of a stream, which is basically an in-memory table; and iv. many existing tools can generate standard SQL.

The STREAM keyword is the main extension in Calcite's streaming SQL. It tells the system that the query target is incoming rows, not existing ones. If omitted, the query is processed as a regular, standard SQL.\footnote{Note that these semantics are different from those for the \texttt{STREAM} keyword proposed in \ref{sec:standard-stream-materialization}.}

\subsection{Streaming SQL in Apache Flink} \label{sec:flink-impl}

Apache Flink \cite{asf2018flink} is a framework for stateful computations over data streams. Flink provides APIs to define processing logic and a distributed processing engine to apply the logic in parallel on data streams. The system aims to cover the full spectrum of stream processing use cases, including processing of bounded and unbounded as well as recorded and real-time data streams. Flink is suitable for batch and streaming ETL and analytics use cases and event-driven applications, i.e., applications that apply sophisticated business logic on individual events. Flink provides two core APIs, the DataStream API and the DataSet API. While the DataSet API is tailored towards processing of bounded data, the DataStream API provides primitives to handle unbounded data.

\subsubsection{Stateful and Event-Time Stream Processing}
Many of Flink's unique features are centered around the management of application state. The state of operators is locally maintained in a pluggable state backend, such as the heap of the JVM or RocksDB \cite{rocksdb2018}. Since all state is kept locally (in memory or on disk), state accesses are fast. To ensure fault-tolerance, Flink periodically writes a consistent checkpoint of the application state to a persistent storage system, such as HDFS or S3 \cite{carbone2015apache}. For recovery, the application is restarted and all operators are initialized with the state of the last completed checkpoint. Based on checkpoints, Flink also supports to stop and resume applications, update application logic, migrate applications to different clusters, and scale them up or down.

Another aspect of stream processing that Flink handles well is event time processing. Following the Millwheel model \cite{akidau2013millwheel}, Flink supports record timestamps and watermarks to decide when a computation can be performed. 

\subsubsection{Relational APIs} Apache Flink provides two relational APIs, the LINQ-style \cite{meijer2006linq} Table API and SQL. For both APIs, Flink leverages Apache Calcite \cite{asf2018calcite} to translate and optimize queries. Note that queries of both APIs are translated into a common logical representation, i.e., a tree of Calcite RelNodes. Depending on whether all base tables of a query are bounded or not, a different set of optimization rules is applied and the query is translated into a Flink DataSet program or into a DataStream program. Regardless of the chosen target API, the semantics of the resulting programs are identical, i.e., their results are the same given the same input data.

The execution of programs generated for bounded input is similar to traditional query processing techniques of database systems or comparable SQL-on-Hadoop solutions. However, DataStream programs generated for relational queries on unbounded and continuously arriving data streams follow the principles of query evaluation on time-varying relations with event time semantics as described earlier in this paper. In the following, we describe Flink's implementation of the model and its current state in Flink 1.7.0.

\subsubsection{Implementation} A query processor that evaluates continuous queries on time-varying relations needs 1) a mechanism to encode and propagate arbitrary changes of input, intermediate, or result relations and 2) implementations for relational operators that consume changing input relations and update their output relation correspondingly.

Apache Flink features two methods to encode changes of a relation, \textit{retraction streams} and \textit{upsert streams}. Retraction streams encode all changes as INSERT or DELETE messages. An update change is encoded by a DELETE message followed by an INSERT message. An upsert stream requires a unique key and encodes all changes as UPSERT or DELETE messages with respect to the unique key. Hence, upsert streams encode UPDATE changes with a single message. While retraction streams are more general because they do not require a unique key, they are less efficient than upsert streams.

For many relational operators, implementations that process time-varying relations need to keep the complete input or an intermediate result as state in order to be able to correctly update their output relation. For example, a join operator fully materializes both input relations in order to be able to emit new result rows or update or delete previously emitted result rows. Operators that process a single row at a time, such as projections or filters, typically do not require state and can simply adjust and forward or filter change messages. 

However, there are several special cases for operators that allow to reduce the amount of state that needs to be maintained to evaluate the operator logic on time-varying relations. These special cases are related to event time attributes for which watermarks are available. Operators, such as GROUP BY aggregations with event time windows, OVER windows with an ORDER BY clause on an event time attribute, and joins with time-windowed predicates can leverage the watermarks to reason about the completeness of their input. 

Flink's implementations for such time-related operators perform a computation when the input for the computation is complete as indicated by the watermarks. The result of the computation is emitted and all state that is no longer required is discarded. Therefore, Flink's materialization strategy is currently fixed to watermark-completeness, i.e., results are emitted when all required input was received and late arriving rows are discarded. Hence, time-based operators will only append new output and never delete or update previously emitted result rows.

Flink's SQL runtime is based on operator implementations that do not keep any state, that fully materialize input or intermediate results, and that automatically clean up their state based on progressing time. Flink chooses operator implementations during optimization. Whenever the optimizer detects an opportunity to evaluate an operator with an implementation that is more space-efficient, it picks that implementation.

An important aspect of leveraging event time and watermarks in SQL is to track whether an operator preserves the alignment of event time attributes with the watermarks or not. The alignment is easily lost if a projection modifies event time attributes or if an operator implementation shuffles the order in which rows are emitted. For now, Flink follows the conservative approach of degrading an event time attribute to a regular \texttt{TIMESTAMP} attribute if it is not verbatim forwarded by an operator or if an operator implementation will not emit its output aligned with the watermarks.

\subsection{Streaming SQL in Apache Beam}
\label{sec:beam-impl}

Apache Beam \cite{asf2018beam} is a semantic model for massively parallel computation with a formalized representation as protocol buffer data structures allowing for portable execution of pipelines across multiple platforms. It presents a unified approach to batch and streaming computation \cite{akidau2015dataflow}, and frames a way of thinking about data processing around answering four questions:

\begin{itemize}
    \item \textit{What} results are being computed?
    \item \textit{Where} in event time are results being computed?
    \item \textit{When} in process time are results materialized?
    \item \textit{How} do refinements of results over time relate?
\end{itemize}

Notably, the Beam model is language-independent and backend-agnostic. Nodes contain user-provided processing logic which is executed by a language-specific coprocessor.

Beam's primary reason for existence is to allow users of any language to execute their big data computations on the backend of their choice. At the time of this writing, a Beam pipeline (depending on features used) can be executed on Apache Flink, Apache Spark, Apache Apex, Apache Samza, Apache Gearpump, and IBM Streams (with prototypes for Apache Tez, Apache Hadoop MapReduce, and JStorm). Libraries exist for directly authoring Beam pipelines using Java, Python, and Go. Many users do not directly author a pipeline, but interact with a higher-level library that generates pipelines, including Spotify's Scio\cite{scio} (in Scala), Seznam.cs's Euphoria\cite{euphoria} library (a more fluent Java library), and of course Beam SQL.

\subsubsection{Beam SQL}

Beam SQL is an implementation of SQL using Beam's primitive transforms. As such, it automatically works for streams!  SQL can be embedded as a transform in a pipeline. Beam also includes an interactive shell for working entirely in SQL. Because it compiles only as far as Beam's high-level primitives, Beam SQL -- as an embedded transform or via the SQL shell -- can be executed without modification on any engine with a Beam runner.\footnote{This includes, amusingly enough, Flink. We leave it as an exercise for the reader to rigorously evaluate the differences in the two SQL variants.}

Beam uses the Calcite parser and optimizer to optimize a query's logical plan into a "physical" plan that can be directly translated into a Beam pipeline. Experience translating SQL to Beam's primitives and applying SQL as a transform in a Beam pipeline exposes some key issues that the proposal in this paper hopes to resolve:

\textbf{Beam does not have adequate built-in support to represent a materialized changelog.} Because changelogs include deletions as well as insertions, they require what is know in Beam as "retractions", a feature proposed but never fully designed or implemented. So Beam SQL carefully limits queries to those that only ever add rows to their output.

\textbf{There is no notion of event time columns in SQL or Calcite.} Instead, one directly writes \texttt{GROUP BY TUBMLE(...)} (for example) in order to provide a grouping key with an upper bound in event time. Notable, \texttt{GROUP BY HOP(...)} violates the relational model by not actually being a grouping of the input rows, but either a multiplication of the rows or a dropping of some rows, depending on parameters. Similarly the relationship of \texttt{GROUP BY SESSION(...)} to usual SQL semantics is unclear, though the same logic can probably be defined as a specialized analytic function.

\textbf{Materialization preference ("triggering") cannot be expressed within SQL.} A user is free to express triggering in their pipeline and Beam SQL will respect it where possible. However, as mentioned above, in the absence of retractions many queries must simply be rejected in the presence of triggers that would cause incorrect results, such as triggers combined with shuffle-based equijoins.

\end{document}